# A Tissue Engineered Model of Aging: Interdependence and Cooperative Effects in Failing Tissues

A. Acun[1], D. C. Vural[2], P. Zorlutuna[1, 3]*

[1]Bioengineering Graduate Program, University of Notre Dame, IN

[2]Department of Physics, University of Notre Dame, IN

[3]Aerospace and Mechanical Engineering, University of Notre Dame, IN.

**\* Corresponding Author:** Pinar Zorlutuna

**Abstract**:

Aging remains a fundamental open problem in modern biology. Although there exist a number of theories on aging on the cellular scale, nearly nothing is known about how microscopic failures cascade to macroscopic failures of tissues, organs and ultimately the organism. The goal of this work is to bridge microscopic cell failure to macroscopic manifestations of aging. We use tissue engineered constructs to control the cellular-level damage and cell-cell distance in individual tissues to establish the role of complex interdependence and interactions between cells in aging tissues. We found that while microscopic mechanisms drive aging, the interdependency between cells plays a major role in tissue death, providing evidence on how cellular aging is connected to its higher systemic consequences.
**Keywords:** Aging, tissue engineering, systemic interdependence

Many simple organisms such as ferns, hydra or jellyfish, do not age (*1*). They have a constant mortality rate due to intrinsic or extrinsic "accidents". In contrast, in complex species, at any given time, an older individual is more likely to die than a young one. Furthermore the functional form of mortality rate $\mu(t)$ exhibits a remarkable cross-species universality from worms and insects to birds and mammals, characterized by a sharp drop early in life, followed by a near-exponential increase (Gompertz Law) followed by a late life plateau (*1-3*). The lack of aging in simple systems, and the similarity of mortality curves in complex organisms suggest that aging is an emergent, systemic property.

Aging is the outcome of a long evolutionary history (*4-6*) and is microscopically driven by well-understood biomolecular processes (*7, 8*). However organisms die not because they run out of cells, but rather due to complex systemic problems. Thus, a complete theory of aging must go beyond cellular damage. In our recent theoretical model we described aging, failure, and death in terms of a cascade of failures taking place on a complex network of interdependent nodes (*9*). The basic assumption in this model is that nodes malfunction when other nodes they rely on also malfunction. As a result some failures cause others, which in turn propagates further, ultimately leading to a singular system-wide catastrophe. In simulations the catastrophe occurs suddenly as soon as the malfunctions accumulate to a critical point. Furthermore it was shown in (*9*) that the collapse curves, the damage and repair rate dependence of network lifetimes, and event size distributions are very weakly dependent on the structure of the interdependence network; and theoretically obtained mortality curves were shown to be in good agreement with that observed in a large variety of complex organisms and machines. In this work we experimentally verify the elements of this theory with the use of synthetic tissues and demonstrate for the first time the systemic spread of failures from cellular level to tissue level, and propose a fresh view of systemic aging.

Consider a hypothetical organism whose organs/tissues/cells are spread sufficiently far apart to prohibit any interaction, but maintained well enough to avoid immediate death. Will this organism exhibit a monotonically increasing probability of death or display age associated problems? According to the interdependence network picture, one should see no systemic effects, nor any cascading failures due to intercellular interactions. In this case, one should see a nearly constant mortality rate $m = \Delta n/n$ only due to aging at the cellular level, thus, an exponentially decaying cell population $n(t) \sim e^{-mt}$. However, when cells are nearby, and thus interacting, we expect to see deviations from an exponential decay of population. More precisely, we expect to see a monotonically increasing mortality rate $\Delta n/n$, which implies an increase of downwards slope of $n(t)$ (for full theoretical details cf. Materials and Methods and (*9*)).

One of the outcomes of this study is to demonstrate the above ideas experimentally, and quantify the interactions between cells that lead to tissue-level failure. While it is not possible to control the cellular level aging characteristics and cell-cell interaction strengths of *in vivo* tissues parametrically, tissue engineering techniques allow for direct tests on synthetic live tissues and organoids that can recapitulate native ones. These structures are sufficiently complex and controllable that we can explore the hypothetical idea discussed above, to shed light on the nature of aging. Synthetic tissues allow controlling damage and repair rates, cell viability, and thus, allow direct observation on how cells influence one other's performance upon failure. Most recently, synthetic tissues were used to investigate various diseases, including age related ones, and successfully established several platforms, such as Alzheimer's disease in-a-dish (*10*) and Barth syndrome on-a-chip (*11*). Therefore, using synthetic tissues, i.e. aging in a dish, to parametrically study a phenomenon as complex as aging, is a highly promising approach.

In this work we study the hierarchical spread of failure from cells to tissues by growing synthetic tissues in well-controlled hydrogel microenvironments in which we vary intercellular distance, environmental stress, and the age of cells. We first establish that systemic aging, in contrast to cellular aging, is real and significant, and that the effect vanishes when well maintained cells are spread far apart. In other words, dense tissues display larger "age-specific mortality", whereas the mortality rate of sparse tissues is near constant. We then determine the relative importance of systemic aging to cell-level aging by comparing the population curves of synthetic tissues made of young cells of varying density (and thus, interaction strength), to that of synthetic tissues made of "aged" cells. Finally, we exchange the culture media of the synthetic tissues made of young and aged cells with different cell population densities, much like the recent parabiosis experiments (*12, 13*), to identify the mechanism behind systemic aging.

**Results and Discussion**

Our findings support that aging cannot be solely explained by failures of individual cells but is an emergent phenomenon involving strong intercellular interactions. Specifically: **(i)** We find that systemic aging is a more important factor than cellular aging (regardless of how aging is induced). A healthy young cell is more likely to die if its neighbors malfunction, than an old or stressed cell with intact neighbors. **(ii)** We find that cellular aging is tightly coupled to systemic aging, since aging in the cellular level causes cells to lose their ability to interact with surrounding cells. Specifically, we determined that one of the causes underlying systemic aging is the loss of ability to receive or internally process functional cooperative factors from surrounding cells, but not a loss of ability to produce cooperative factors, or a loss of function of the produced cooperative factors.

Our aged tissue model consists of neonatal rat cells treated to exhibit senescence markers, and a synthetic polymer, poly(ethylene glycol) (PEG) 4-arm acrylate modified with cell attachment peptide arginine-glycine-aspartic acid (RGD) (fig. S1A), that provides a controlled, biomimetic 3-D microenvironment (*14-17*). RGD-modified PEG (PEG-RGD) allows for cells to attach and spread similar to native tissue structure, while it prevents them from dividing or migrating since it is not enzymatically degradable by the cells. This allows us to control the localization, hence the distance between the individual cells constituting the tissue In addition, the PEG-RGD hydrogels provide a stiffness of around 10 kPa for all cell encapsulation densities used (fig. S1B). This provides a physiologically relevant system for the cell type used in this study (i.e. CFs), as the native heart muscle stiffness is 10 kPa at the beginning of the diastole (*18*). In addition, the synthetic tissues with different encapsulation densities did not show any significant difference in their stiffness, ruling out any possibility of mechanical microenvironment contributing to the observed differences in cell survival. In order to mimic cellular level senescence, we artificially aged primary neonatal rat cardiac fibroblasts (CFs) through applying different kinds of cell-level stresses, which we refer to as "pre-aging" conditions. This way we aimed to test the effect of different types of cellular level damage on systemic tissue level failure. Two of the most important hallmarks of aging are cellular senescence and genomic instability (*19, 20*). Importantly, cellular senescence and DNA damage are not only observed in in vitro culture systems but also observed in aged tissues of various animal models (*21*) and humans (*22, 23*).

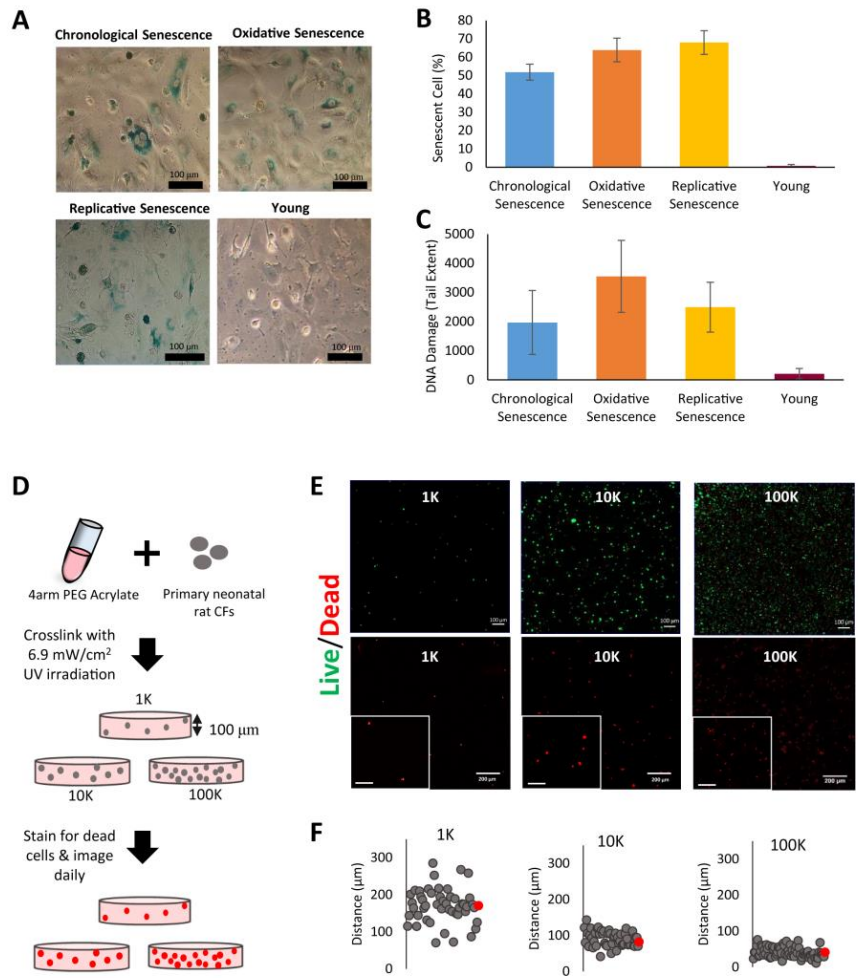

**Fig. 1 The preparation and characterization of pre-aged cells and fabrication of respective synthetic tissues.** (A) Senescence marker staining, (B) the senescent cell percent, and (C) DNA damage in pre-aged and young cells. (D) The schematic of fabrication and staining, and (E) the day 1 live/dead (top row) and dead cell staining (bottom row) (dead, red=ethidium homodimer-1; live, green= calcein-AM) images of 1K (day 1), 10K (day 6), and 100K (day 8) chronologically aged tissue constructs.(F) The cell-cell distance distributions, of the chronologically aged synthetic tissues with 1K, 10K, and 100K population densities (red dot represents the average cell-cell distance). (Scale bars=100 µm for live/dead images and inset images, and 200 µm for dead cell images).

Therefore, to construct physiologically relevant aged tissue models, we induced cellular senescence and DNA damage in CFs either by restricting the cells to a defined attachment area for a month (Chronological Senescence), applying a low dose of oxidative stress in addition to restricting to a defined attachment area for a month (Oxidative Senescence), or by having cells reproduce multiple times until they reach their replicative limit (Replicative Senescence). Several other studies have shown that treatments similar to the pre-aging conditions used in this study induce senescence and DNA damage in vitro: Replicative senescence is induced through multiple population doublings in various cell types *(24-26)*; senescence associated with oxidative stress was exemplified in literature *(27, 28)*; and the chronological aging in a skin tissue equivalent has been shown in vitro *(29)*. As control, we prepared tissues using the cells isolated from young animals and simply passaged twice in regular cell culture conditions (Young). At the end of the pre-aging period, we confirmed that the cells showed the characteristic symptoms of aging for all three pre-aged groups.

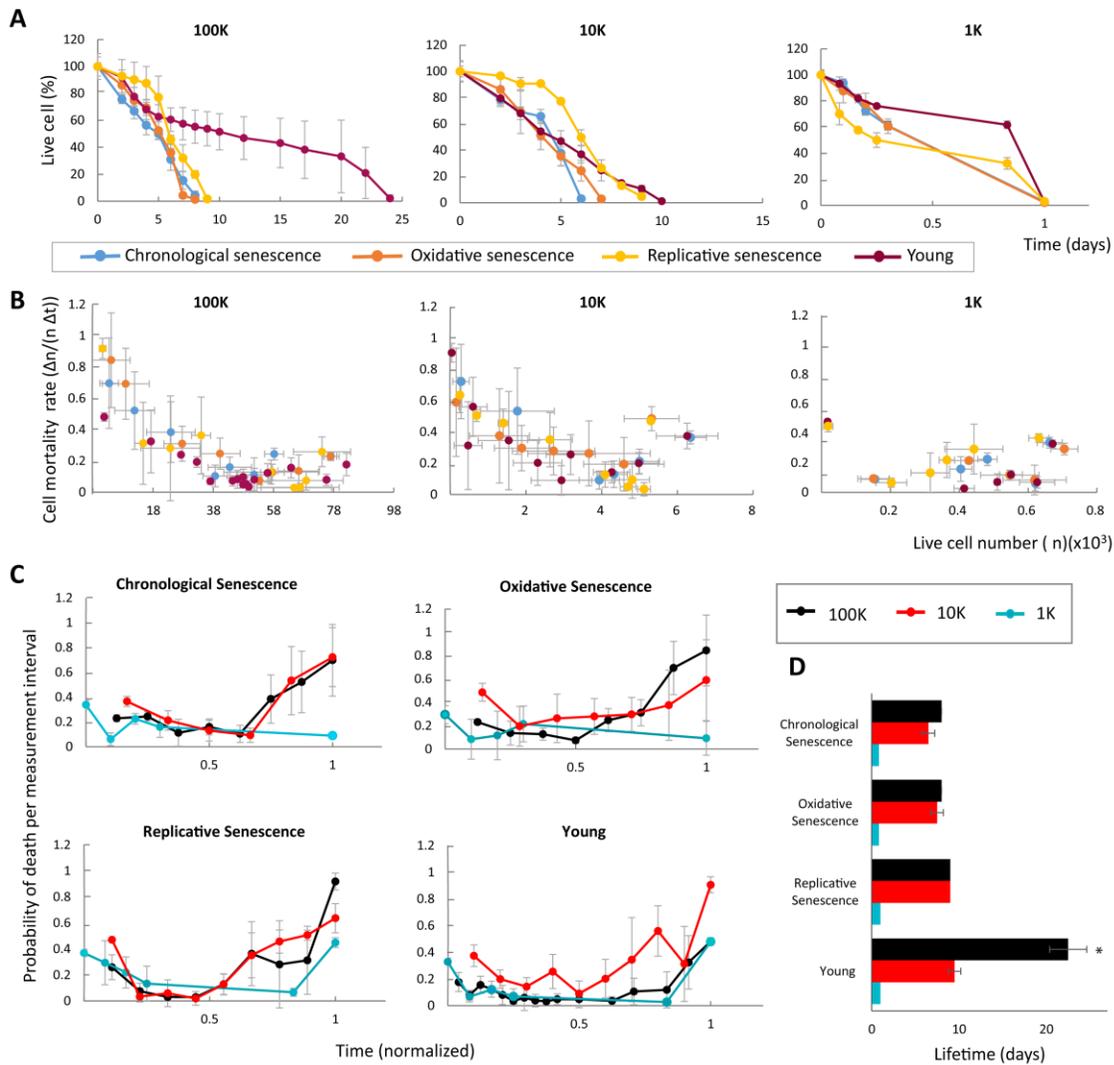

**Fig. 2 Failure characteristics of synthetic tissues made from young or pre-aged cells**. The change of (A) live cell percentage over time, (B) cell mortality rate with respect to live cell number, and (C) probability of death over time. (D) The lifetime of synthetic tissues made from young or pre-aged cells with different population densities (* indicates significance, p<0.05). Denser tissues where cells can interact, display age specific mortality, whereas the mortality rate of sparse tissues is closer to constant.

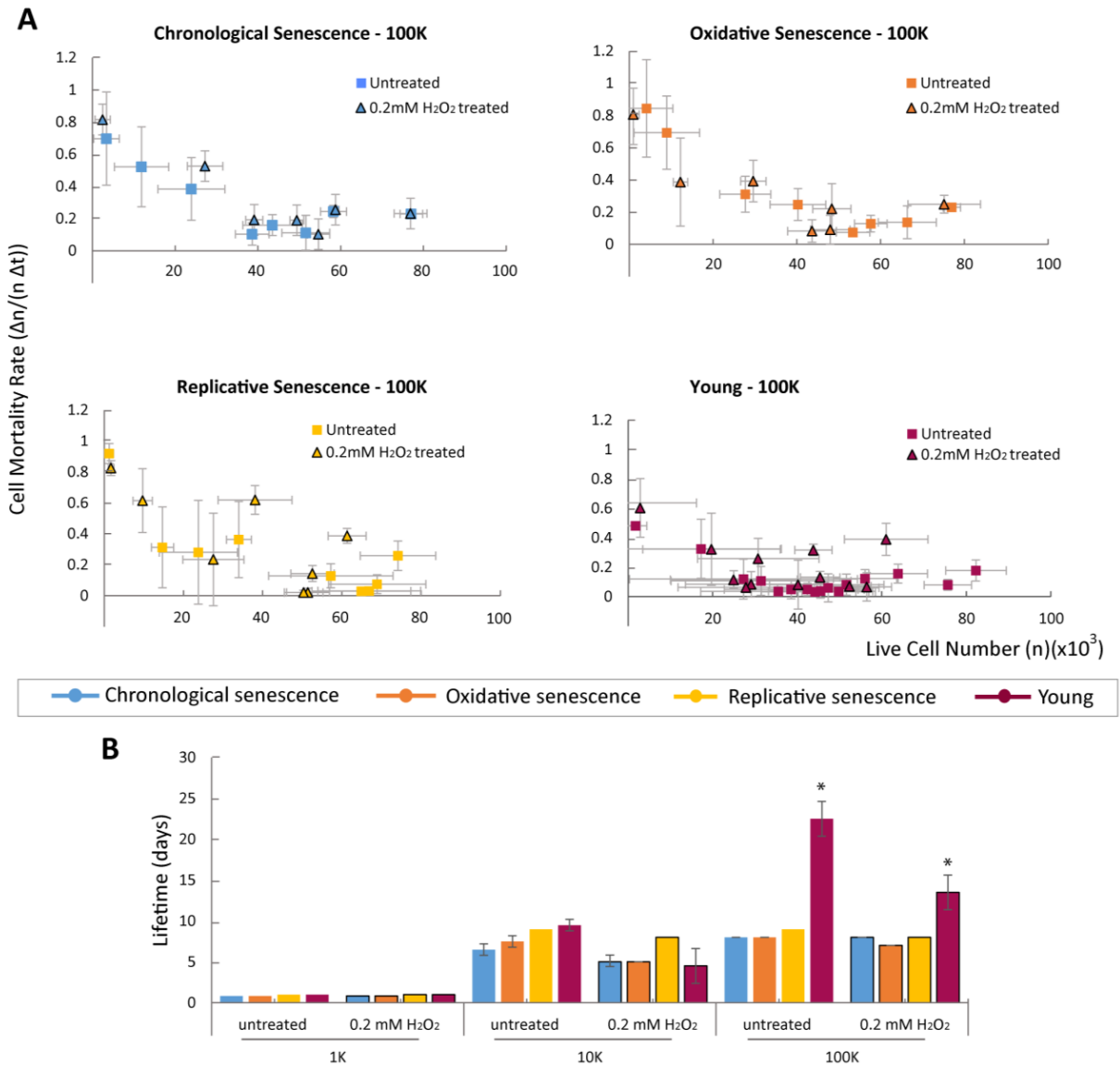

**Fig. 3 Determining the resilience of synthetic tissues made from young or pre-aged cells against oxidative stress and the effect of different types of cellular level damage.** (A) The change in cell mortality rates of 100K tissues with respect to the change in population density. (B) The lifetime of synthetic tissues made from young or pre-aged cells with or without oxidative stress (* indicates significance, $p<0.05$).

The pre-aging conditions resulted in cellular senescence in more than 50% of the population, whereas less than 1% of the young cells showed senescence as shown through senescence associated β-galactosidase staining (Fig. 1A and B). We also tested the cyclin-dependent kinase inhibitor 1 (p21) expression of the cells pre-aged using the 3 abovementioned techniques (fig. S2A). p21 expression is known to be elevated in aged cells and tissues *(30)*, thus is used as an aging marker. Therefore, the increased expression of p21 in the pre-aged cells further confirmed the aged phenotype of the cells. In addition, the characteristic enlarged and flattened morphology of replicative senescent cells *(31)* was observed in our "replicative senescence" pre-aged group (fig. S2B).

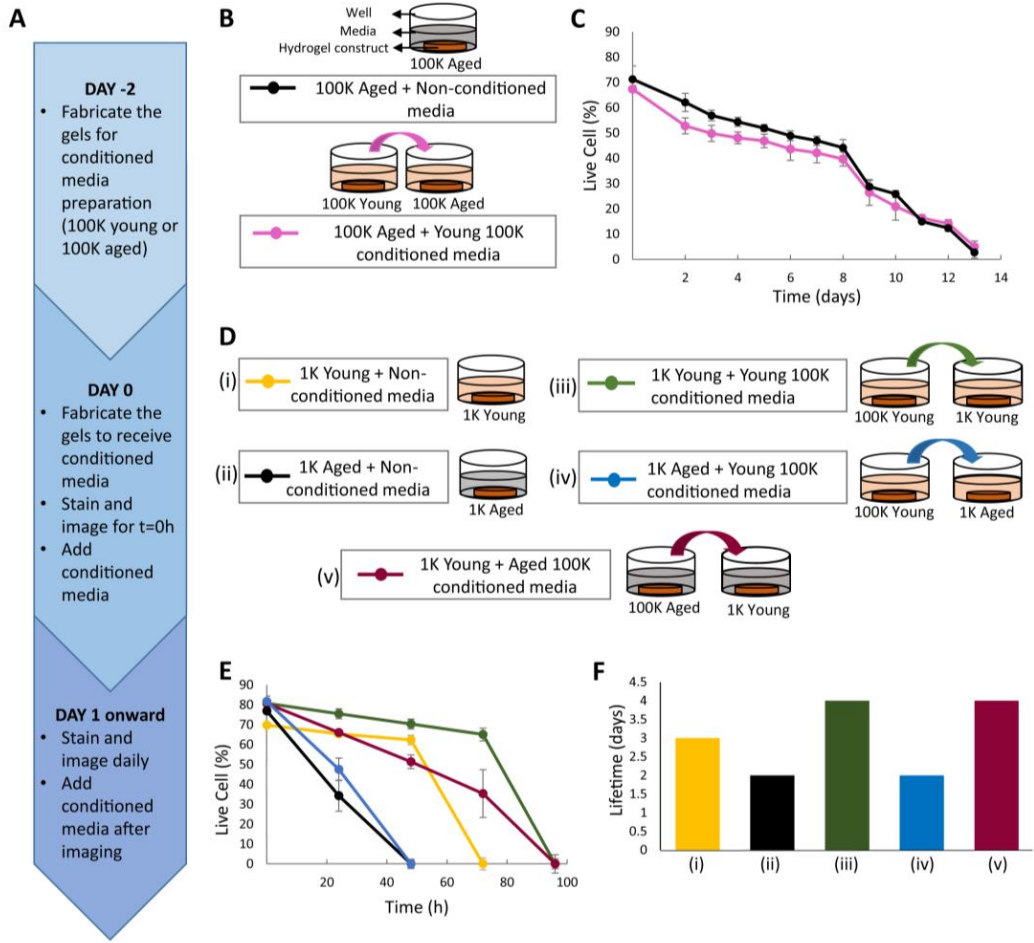

**Fig. 4 Determining the effect of secreted factors on failure of synthetic tissues made from young or pre-aged cells.** (A) The timeline for conditioned media experiments. (B) The schematic of media transfer and (C) the live cell percentage change over time of synthetic tissues with 100K pre-aged cells receiving 100K young synthetic tissue conditioned or non-conditioned media. (D) The schematic of media transfer experiment of synthetic tissues made from young or pre-aged cells at 1K density, receiving 100K young or 100K pre-aged synthetic tissue conditioned media. Synthetic tissues maintained in non-conditioned media are used as controls. (E) The live cell percentages of 1K synthetic tissues receiving non-conditioned (control) or conditioned media. (F) The lifetime of synthetic tissues made from young or pre-aged cells at 1K density receiving 100K young or 100K pre-aged synthetic tissue conditioned or non-conditioned media.

We also determined the DNA damage using comet assay in pre-aged and young cells, and observed that all pre-aged groups showed a significantly higher level of DNA damage ($p<0.05$) (Fig. 1C) further confirming the success of our artificial aging treatment in presenting symptoms that are associated with aging on cellular level. Once the pre-aging treatment was completed the cells were encapsulated in PEG-RGD hydrogels at different densities (Fig. 1D) (fig. S3).

To tune the interactions between cells we varied the cell density. We controlled intercellular distance by using 3 different encapsulation densities: $1 \times 10^5$ (100K), $1 \times 10^4$ (10K), $1 \times 10^3$ (1K) per construct. These 3 densities were chosen to achieve different degrees of intercellular communication through different degrees of secreted factors among the cells, representing dependent (100K) and non-dependent (1K) populations, respectively. In addition, by using a synthetic hydrogel without enzymatic sequences that can be degraded by the cells, we made sure that this density is maintained throughout the course of the experiment, as the cells can neither move nor proliferate through the material. The synthetic tissues were then cultured in normal cell culture conditions, and stained for dead cells using ethidium homodimer-1 and imaged daily. The number of dead cells was counted daily using imageJ software and staining/imaging/counting continued until at least 98% of the population was recorded to be dead (Fig. 1E). At this point the tissues were assumed to be dead and the day that this ratio was reached was recorded as the lifetime of the samples.

We estimated the cell-cell distance by considering the volume of the whole construct and the total number of cells, assuming the cells were uniformly distributed throughout the gels. We also experimentally determined the probability distribution for nearest neighbor distance of cells for the 3 densities used (Fig. 1F). We observed that the estimated cell-cell distance (170 µm, 80 µm, and 35 µm for 1K, 10K, and 100K, respectively) was achieved in the synthetic tissues with cell-cell distances averaging at 170.1±46.7 µm, 90.7±20.9 µm, and 41.1±11.5 µm for 1K, 10K, and 100K, respectively.

We observed that the synthetic tissues with pre-aged cells died sooner than the synthetic tissues with young cells only when the population was dense enough to interact: for the dense, 100K, tissues, ones with young cells lived 25 days, while the tissues with pre-aged cells lived 8-10 days. As the population density decreased, the difference between the tissues composed of pre-aged and young cells disappeared (Fig. 2A). This result suggests that the complex cell-cell interactions provided in a dense population have an effect on aging through cellular interdependence. In addition, aging at the cellular level prohibits these interactions only when the cell population is dense enough to interact while the kind of cellular damage makes little difference.

We also calculated the cell mortality rate and determined how it changes as a function of cell density (and thus, strength of intercellular interactions). We found that cell mortality rate increases as the populations get sparser in 100K and 10K tissues (Fig. 2B). Interestingly, for populations lower than 1K the cells are sufficiently spread apart that systemic aging completely vanishes, and the mortality rate is nearly constant. Denser populations died at an increasing rate as the live cell number decreased due to the cascading failures propagating across the interdependent cell network. In contrast, our sparse groups displayed very little signs of systemic aging, regardless of how the cells were pre-aged in advance.

In addition, as functional cells become sparser over time, both pre-aged and young cells were able to interact less and therefore died at an accelerating rate, which suggests that the disruption of cell-cell interactions in an interdependent population is the essential cause of aging. In other words, systemic aging can happen even without cellular level aging. Thus, individual cell damage must only be a proximal cause of aging.

In a realistic biological setting, an increase in cell failure rate over time logically implies an increase in tissue/organ mortality rate over time, in the demographic sense. Thus, our observation that larger interactions cause larger increases in mortality rates is in qualitative agreement with how simple organisms (with decentralized interdependence structures) have a near-constant mortality rate while complex organisms (with tightly knit long range interdependence structures) tend to have rapidly increasing mortality rates.

To see the systemic effects clearly, we plot the cell death rate in tissues with different population densities against normalized time (Fig. 2C). While denser populations live longer, they also display stronger aging features, as measured by the increase in cell mortality rate over time. The sparsest group, 1K, displays weaker signs of aging, indicating that cell deaths are solely due to damage in individual cells but not due to complex interdependence of the entire system. This is further supported by the longer lifetimes of young tissues only at the highest population density, or where the most complex cell-cell interactions are expected to be established (Fig. 2D), whereas the lifetime of all groups was approximately the same at both 10K and 1K population densities, regardless the cells were pre-aged or not, and regardless which way the cells were pre-aged.

Another important question about aging is whether aged cells/tissues respond to stress different than young cells/tissues. Accumulation of oxidative stress has been proposed to be a major factor in aging (*32-34*). It is also present in tissues in high amounts under life threatening conditions such as myocardial ischemia. Thus, we investigated how different pre-aging conditions affect resilience to oxidative stress (Fig. 3). We exposed the synthetic tissues made from young or pre-aged cells to oxidative stress (0.2 mM $H_2O_2$) throughout the lifetime measurement experiments and determined their death characteristics. We observed that having functional neighbors is more important than being stress-free, as the young cells survived the oxidative stress better only when the population was dense. The ability of the young tissues to cope with stress decreased as the population density decreased, further suggesting that the complex interdependence formed at higher densities is the key for how tissues fail regardless of environmental stress (fig. S4).

Similarly, the probability of death of sparse 1K tissues remained constant while that of 10K and 100K increased in time, following the same pattern as the tissues that were not exposed to any post-stress (fig. S5). The cell mortality rates followed a similar pattern whether stress was applied or not, among the same pre-aging condition groups (Fig. 3A). The 10K and 1K populations of the pre-aged and young groups also followed the same pattern for cell mortality rate, regardless of the stress applied (fig. S6 and S7). In addition, when the population was sparse, tissues with young cells coped with stress equally worse as tissues with pre-aged cells due to their lack of interactions. When the population was dense, however, tissues with young cells coped with stress significantly better than tissues with pre-aged cells. This indicates that cellular level aging decreases stress resilience by disrupting cellular interactions (Fig. 3B).

We further investigated the interaction between cellular level aging and systemic aging. Specifically, we determined how cellular level aging disrupts the interactions to cause the tissues to collapse sooner and how they show a lower resilience to stress. We considered two possibilities: (i) Aged cells lose their ability to produce (functional) cooperative factors. (ii) Aged cells do produce cooperative factors, but they cannot make use of them due to degrading surface receptors or degrading downstream pathways. As discussed below, we find evidence to suggest that hypothesis (ii) is more plausible.

To test between these two hypotheses we designed an experiment where we transferred the conditioned media from high (100K) population density tissues with young cells to tissues with pre-aged cells (Chronological Senescence) with high (100K) population density (Fig. 4A and B). We observed that the aged cells did not live longer when they received the young conditioned media. Their population curve followed almost the same profile as the control (tissues with aged cells that did not receive young media).

This suggests either that the tissues with aged cells are not able to receive or process cooperative factors or that the factors are received and processed but their quantity is insufficient for the dense aged tissue. We eliminated the latter option by performing another experiment in which we transfer the conditioned media of 100K tissues with young/aged cells into 1K tissues with young/aged cells. If the young conditioned media was helpful to the aged tissue but was quantitatively insufficient, it should cause the aged 1K tissues to live longer compared to controls receiving non-conditioned media (Fig. 4C). If the aged cells are still producing the required factors, then the conditioned media from aged 100K cells should aid young 1K cells to live longer, as 100K populations live longer than 1K populations, regardless of age (Fig. 2D).

Our results showed that the aged 1K tissues died sooner than all young 1K tissues, and receiving young 100K conditioned media did not improve their lifetime (Fig. 4D). This suggests that the aged cells cannot receive/process cooperative factors even when factors are functional and sufficiently available suggesting that the aged tissues show a functional loss through the pre-aging period.

We verified the availability of functional factors by the improved survival of young 1K tissues when they received the conditioned media from synthetic tissues made from young or pre-aged cells at 100K density. This also suggests that the aged cells do produce cooperative factors that help each other survive, however, their ability to receive/process them is compromised. Furthermore, when we added oxidative stress to the conditioned media and non-conditioned media, neither the lifetime nor the aging profile of the synthetic tissues made from young or pre-aged cells were affected (fig. S8).

Presently available aging research primarily focuses on the role of various cellular and bio-molecular mechanisms on cell damage (*35-37*). Our experimental evidence supports the thesis in (*9*) that a viable theory of aging must take into account the *emergent nature* of aging, and describe it as a universal property of a strongly interdependent collection of failure prone components. While aging is clearly driven by microscopic mechanisms, it is "more than the sum" of these microscopic factors. Grounded by empirical evidence, we have directly and quantitatively observed how the interdependency between cells play a role in tissue death, and bridged cellular aging to its higher systemic manifestations. The molecular mechanisms behind the functional loss in aged tissues we observed in this study need to be explored further including the differences in cooperative factors produced by young versus aged tissues. We aim to investigate the differences in media components of aged and young tissues under normal and stress conditions in our future studies.

Generally speaking, the picture of failure-prone interdependent components is not specific to multicellular life: some of the ideas presented here may also potentially provide further insight into the aging of mechanical, social, political and ecological systems.

**Materials and Methods**

All animal experiments were performed using protocols approved by Institutional Animal Care and Use Committee (IACUC) of University of Notre Dame, in accordance to the guidelines of National Institutes of Health, Office of Laboratory Animal Welfare.

Cell Culture

Neonatal rat cardiac-fibroblasts (CFs) were isolated from 2-day-old Sprague-Dawley rats (Charles River Laboratories). The isolation was carried out following a protocol approved by the guidelines of the University of Notre Dame, which has an approved Assurance of Compliance on file with the National Institutes of Health, Office of Laboratory Animal Welfare. Briefly, the rats were sacrificed by decapitation after $CO_2$ treatment, and the hearts were immediately excised. CFs were isolated using a previously established protocol *(38)*. Briefly, the rat hearts were minced and incubated in trypsin (Life Technologies) (4°C, 16h) with gentle agitation. Next, the heart pieces were further digested by adding collagenase type II (Worthington-Biochem) (37°C) and agitating several times with subsequent trituration in order to digest the extracellular matrix. The undigested tissue pieces were filtered and the filtrate was plated (37°C, 2h). The filtrate at this point contained CFs and cardiac myocytes (CMs), which were separated making use of their differential attachment during a 2h plating period. The CFs attach to the culture plate at the end of this 2h where CMs remain in solution. Next, the CMs in the solution were removed from the plate and DMEM supplemented with 10% fetal bovine serum (FBS) and 1% penicillin/streptomycin (p/s) (standard culture media) was added to CFs for maintaining the culture. At this point the CFs were labeled as passage 1 (P1) and media was changed every 3 days. CF cultures were passaged at approximately 80% confluency using trypsin-EDTA (0.05%) (Life Technologies) and maintained in standard culture media until the start of pre-aging treatments.

Pre-aging of CFs

CFs were aged in culture prior to synthetic tissue fabrication in 3 different scenarios: (1) CFs (P3) were maintained in 100% confluency without passaging for 30 days in standard culture media, with media changes every 3 days (Chronological Senescence); (2) CFs (P3) were maintained in 100% confluency without passaging for 30 days under low oxidative stress conditions (standard culture media supplemented with 20 μM $H_2O_2$) with media changes every 3 days (Oxidative Senescence); and (3) CFs (P3) that were allowed to go through approximately 20 population doublings via passaging several times until they reach replicative senescence (Replicative Senescence). CFs (P3) that were not pre-aged were used as controls (Young).

Senescence Marker Staining and Comet Assay

At the end of the pre-aging periods, the CFs were collected using trypsin-EDTA and reseeded to glass cover slips at a density of $1.5 \times 10^6$ cells/mL (n=3). Cellular senescence was determined through senescence-associated β-galactosidase staining using the Cellular Senescence Assay Kit (Millipore) following manufacturer's instructions. The senescent cell percent was determined by counting the total cell number and the number of cells positively stained for senescence-associated β-galactosidase using imageJ software. At least 4 images from each sample from at least 3 different samples were counted and averaged to yield average percentage values.

At the end of the pre-aging period the DNA damage was determined by using comet assay (Cell Biolabs) following the manufacturer's instructions. The assay uses the extent of DNA smear (tail) upon single cell electrophoresis to examine the amount of DNA damage in the cells. Briefly following the agarose gel electrophoresis, the cells in the gels were stained with a DNA dye and imaged using a fluorescence microscope (Zeiss, Hamamatsu ORCA flash 4.0). Then the images were used to measure the tail length using imageJ software. Since the DNA damage is correlated with the DNA tail extent, the average tail extent for different groups was calculated to compare the amount of DNA damage upon various pre-aging conditions performed in this study.

Passage 3 CFs (Young group) were used as control in both assays.

PEG-RGD Hydrogel Preparation and Characterization

The YRGDS (H-tyr-Arg-Gly-Asp-Ser-OH, Mw 596.60, Bachem) sequence was conjugated to acryloyl-PEG-N-hydroxysuccinimide (acryl-PEG-NHS, Mw 3400, Jen Kem Technology) following a previously described protocol *(39)*. The RGD conjugation to PEG was confirmed through Nuclear Magnetic Resonance (H-NMR) – Spectroscopy. RGD conjugated PEG-NHS was mixed with 4-arm PEG-acrylate (Jen Kem Technology) in a 1.5:8.5 (w/w) ratio to prepare PEG-RGD hydrogels. The stiffness of the hydrogels with 3 different cell encapsulation densities ($1\times10^5$, $1\times10^4$, and $1\times10^3$) was determined using a nanoindenter (Piuma Chiaro, Optics11, Amsterdam, The Netherlands) with an indentation probe (spring constant of 0.261 N/m, tip diameter of 16 µm). Elastic moduli of hydrogels were identified using a custom made MATLAB code.

Synthetic Tissue Preparation

Pre-aged CFs were collected at the end of the pre-aging periods, and young cells were collected when they reach approximately 80% confluency, using trypsin-EDTA and encapsulated in PEG-RGD hydrogels (n=6). 3 different cell densities were used for encapsulation to control cell-cell distance in the synthetic tissues: $1\times10^5$ (100K), $1\times10^4$ (10K), and $1\times10^3$ (1K) CFs per construct (5µL total volume). The hydrogel solution was prepared by dissolving PEG-RGD (20% w/v) in PBS and photoinitiator (PI) (Irgacure 2959, BASF) (0.1% w/v in PBS) was added to achieve crosslinking. Next, the resulting solution was mixed with the respective cell suspension at a ratio of 1:1 (PEG-RGD:cell solution) and a final concentration of 10% PEG-RGD and 0.05% PI was achieved. The homogenous cell distribution within the hydrogel was ensured by thorough mixing of the hydrogel solution and cell suspension by gentle pipetting. The mixture was then sandwiched between 100 µm thick spacers using a glass slide and exposed to 6.9 mW/cm$^2$ UV irradiation for 20 sec. This dose of UV and photoinitiator is in the range that have been shown to be safe for cell encapsulation studies *(40)*. The synthetic tissues were washed once with PBS (1-2 mins) and 3 times with fresh standard culture media (15 mins each) right after crosslinking in order to get rid of excess PI.

Conditioned Media Experiments:

For the media transfer experiments, synthetic tissues made from young or pre-aged cells with 100K cell density in each tissue were prepared as previously described. The synthetic tissues were maintained for 2 days prior to transferring their media to other synthetic tissues. On day 3 the synthetic tissues to receive the conditioned media were fabricated and following the initial washes they received the respective conditioned media. Controls were prepared for each condition where the same groups with the same population density received non-conditioned media at the same time points. Both 100K and 1K samples receiving the conditioned media were stained for dead cells and imaged after 24 hours of receiving the conditioned media. The samples were stained/imaged every 24 hours and the number of dead cells was determined for each group daily, as described in the following section.

Determining Cell-Cell Distance Distribution

The distance from each cell to its closest neighbor was determined using imageJ software for 1K, 10K, and 100K tissue constructs. When they reached their lifetime, the constructs were imaged using a confocal microscope by taking z-stacks from 5 different points of each construct. The cell-cell distance are measured from each individual z-slice in order to determine the distance between the cells on the same plane.

Determining Live Cell Percentages and Death Rate of the Synthetic Tissues

Ethidium homodimer 1 (EthD-1) (Life Technologies), a nuclear stain that only stains the dead cells, was used to determine the real time live cell percentages and the death rate of the synthetic tissues, following manufacturer's instructions. After fabrication and the initial washes, the 10K and 100K synthetic tissues were incubated in normoxic or 0.2 mM $H_2O_2$ containing standard culture media for 24h. At the end of 24h the synthetic tissues were washed off their media, stained with EthD-1 and imaged using a fluorescence microscope (Zeiss, Hamamatsu ORCA flash 4.0). Following imaging the tissues were washed off the stain and the respective media was added. This procedure was repeated every 24h until the live cell percentage was at most 2%. 1K synthetic tissues were stained with EthD-1 following the same procedure, right after the fabrication and initial washes to determine the t=0h viability. Then the staining was repeated at 2h, 4h, 6h, 20h and 24h, unless stated otherwise.

At the end of each staining the imaging was done by taking z-stacks of the gels at 3 different points, covering approximately 1/8 of the whole gel. Through structural illumination (Apotome, Zeiss) we were able to image our tissues with optical sectioning and eliminated any signal coming from the above or below planes. This allowed us to determine the number of dead cells in each plane separately. For 10K and 100K synthetic tissues the number of dead cells in the 3 middle slices of each stack were counted separately (in total 9 images per tissue) and averaged. This average was then multiplied by the total number of slices present in a stack (10-11 slices in each stack) and by the ratio of the field of view of one stack to the area of the whole gel (1:8) to yield the average dead cell number per gel. For 1K synthetic tissues the same procedure was used for imaging, however, the maximum projections for each stack were used to calculate the average dead cell number per gel. We used the maximum projections of the stacks as the sparsity of the cells at that density allowed us to count the dead cells in all planes at once, thus eliminating the error coming from assuming a constant cell number in each slice of a stack. Then the counted number of dead cells was multiplied by the ratio of the field of view of one stack to the area of the whole gel to yield the average dead cell number per gel. The dead cell numbers were normalized using a calibration curve (fig. S9). The live cell number on day 0 was assumed to be equal to the theoretical number for the respective population density: i.e. the number of live cells in a 100K tissue on day 0 was taken as $1 \times 10^5$ cells per construct. The average dead cell number calculated for each synthetic tissue was then subtracted from the theoretical total cell number in each construct e.g. for a 100K synthetic tissue, live cell number on day 1 was calculated as: 100,000 – (calculated dead cell number on day 1). Then the live cell number on day 2 was calculated by subtracting the calculated dead cell number from the live cell number on day 1. The live and dead cell numbers for each synthetic tissue were determined at the end of every time point. Then we calculated the percent live cell every day using the theoretical total cell number and the calculated live cell number: Live cell (%) = [(live cell number)/(total cell number)]*100.

In addition, we calculated the mortality rate of cells ($m$) in different synthetic tissues. For a tissue with non-interacting cells, $m$ would be constant. Thus, the probability of death per unit time $-\left(\frac{\Delta n}{n}\right)\left(\frac{1}{\Delta t}\right) = m$ would yield an exponentially decreasing population, $n(t) = n_0 e^{-mt}$. Thus, by measuring the deviations of $n(t)$ from this null-hypothesis, we quantified the strength of interactions between cells. Specifically, we determined how $m = -\frac{1}{n}\frac{dn}{dt}$ depends on $n$ and $t$ separately. If $m(n)$ is an increasing function, this indicates that the cells compete for resources, and would have a higher chance of survival when their population density is lower. If $m(n)$ is a decreasing function, this would indicate cooperativity between cells, i.e. that cells have a higher chance of survival if they are densely packed.

Statistical Analysis
All experiments were conducted with n=3 and the individual experiments were performed twice. All results were represented as mean ± standard deviation. Student's t-test was used for comparing two individual groups. All *p* values reported were two-sided, and statistical significance was defined as *p*<0.05.

Data Availability
All data generated or analyzed during this study are included in this published article (and its Supplementary Information files).

**Acknowledgments:** This study is supported by National Science Foundation (NSF) Award PHY-1607643.

**Author contributions:** A.A. performed the experiments, prepared figures 1-4, wrote the materials and methods and results sections, and helped writing discussion section. D.C.V. and P.Z. designed and supervised the experiments and the data analysis, and wrote the manuscript.

**Additional information**

**Competing financial interests:** The authors declare no competing financial interests.